\documentclass[preprints,article,accept,pdftex,moreauthors]{Definitions/mdpi} 
\firstpage{1} 
\pubvolume{1}
\issuenum{1}
\articlenumber{0}
\pubyear{2026}
\copyrightyear{2026}
\datereceived{ } 
\daterevised{ } % Comment out if no revised date
\dateaccepted{ } 
\datepublished{ } 
\Title{External Entropy Production and Human Evolution toward Multi-body Life}

\Author{Yasuji Sawada $^{1,2}$*\orcidA{} and Kenji Toma $^{1,2,3}$\orcidB{}}

\AuthorNames{Yasuji Sawada and Kenji Toma}

\address{%
$^{1}$ \quad Division for Interdisciplinary Advanced Research and Education, Tohoku University, Sendai 980-8578, Japan\\
$^{2}$ \quad Frontier Research Institute for Interdisciplinary Sciences, Tohoku University, Sendai 980-8578, Japan\\
$^{3}$ \quad Astronomical Institute, Graduate School of Science, Tohoku University, Sendai 980-8578, Japan
}

\corres{Correspondence: yasuji.sawada.a4@tohoku.ac.jp}

\abstract{
Ancient human beings started "external entropy production" in a late stage of evolution, in addition to the internal entropy production by which energy was dissipated within the body of life, as previously described consistently with the birth of life by maximum entropy production principle. In this paper, the mechanism for development of external entropy production, which is strongly related with use of tools and controlling fire, is theoretically investigated. Archaeological data show that the brain size of ancient human beings started rapid increase around 2.5 million years ago when the usage of tools and control of fire started. It may be natural to assume that the rapid growth of brain size is related to the growth of awareness which helped cooperation with the other human beings for control of fire. Coupled equations for the growth rate of brain including awareness and for growth rate of size of the interacting human beings are analyzed. The external entropy production per one human being which is directly related to the group size of cooperating human beings is estimated to increase as $\sim 20/$million years in the beginning from the critical time. This evolution created coexistence of internal entropy production of traditional multi-cellular life and new external entropy production of multi-body life. A psychological problem due to the coexistence of two kinds of entropy production mechanism in human being and concept of technologies based on the present thermodynamic evolution theory are discussed. It is suggested that the evolutionary understanding of the origin of global warming based on the external entropy production may be important to create an useful countermeasure.
}

\keyword{external entropy production; awareness; non-equilibrium thermodynamics; maximum entropy production principle; evolution to multi-body lives} 

\begin{document}

%%%%%%%%%%%%%%%%%%%%%%%%%%%%%%%%%%%%%%%%%%
%\setcounter{section}{-1} %% Remove this when starting to work on the template.
\section{Introduction}

The essential role of maximum entropy production principle (MEPP) for the birth and evolution of life was discussed in the review paper \cite{sawada25}, in which a new concept of "external entropy production" was introduced in relation to the later evolution of life, and preliminary discussion was made for the present global warming problem caused by the external entropy production of homo-sapiens. The definition of external entropy production is the entropy produced outside the body of life, and contrasted with the internal entropy production which is entropy produced within the body of the life.\footnote{The word "external entropy production" is found in \cite{zeraati12}, but with
different meaning.} The importance of this concept is clear, because the quantity of external entropy production of homo-sapiens is more than two thousand times greater than their internal entropy production and still increasing rapidly at present. 

Thermodynamically, nature had selected evolution of the brain to initiate and increase the function for the external entropy production, as far as the thermodynamic condition for an open system far from equilibrium (FFE) is satisfied. The largest energy dissipation in the present world is mass production business and transportation business in addition to domestic air conditioning. The last one had started first and followed by former two later. In this paper, the mechanism of birth of external entropy production is investigated together with the thermodynamic evolution theory based on the MEPP. 

%In the review paper \cite{sawada25}, birth and evolution of life were consistently explained by the MEPP for a local system far from equilibrium. It was only suggested there that evolution proceeded to later stage where homo-sapience started bifurcation of entropy production into two kinds; external entropy production in addition to the conventional internal entropy production within the biological body(2). It may be important to understand how this external entropy production was realized, because the amount of entropy production of this mode is much larger even now and has tendency to increase further compared to the internal entropy production, and therefore threatening human life by global warming. 

The role of MEPP for the birth and evolution was discussed by many researchers including the present authors \cite{martyushev21}. The relation of this principle with the second law of thermodynamics was confirmed by statistical mechanics. The example for this proof is referred to \cite{england13}. This paper has shown $Pt(\sigma)/Pt(-\sigma) \sim \exp(\sigma t)$ from the second law of thermodynamics, where $\sigma = Q/Tt$ is the entropy production, and $t$ is the interval of time over which the system exchanges the heat $Q$, and $Pt(\sigma)$ is a time-dependent probability distribution in steady states. This equation implies that the probability ratio of an event producing entropy production $\sigma$ to the probability of an even producing entropy production $-\sigma$ is greater than unity and increases with $\sigma$. This is equivalent to MEPP when there are several choice of events. This discussion is a natural derivation from second law of thermodynamics, and supports the MEPP for the system FFE. This principle was experimentally verified in material science \cite{malkus58,ben90,hill90}.

%%%%%%%%%%%%%%%%%%%%%%%%%%%%%%%%%%%%%%%%%%
\section{The growth of personal brain and size of society}

\subsection{Birth of signaling cells and neurons supporting multicellular life}

The relation of nervous system and multi-cell bio-system during evolution is not well-known. The evolutionary origin of the nervous system has been a matter of long-standing debate \cite{arendt21}. Although it is not clear at present how neuronal systems started functioning among primitive cell system, it seems more legitimate to ask why cells needed mutual communication. As demonstrated in the review paper \cite{sawada25}, the history of evolution is not necessarily continuation of thermodynamic condition FFE. Continued survival of lives depended on the capability to adapt the hard situation. Some examples of adaptation of life to the change of environment were described in the preceding paper \cite{sawada25}. They could manage to decrease entropy production for survival, corresponding to minimum entropy production \cite{glansdorff71}.

A related behavior is found in dictyostelium discoideum (D. D.) \cite{dusenberry96}. One of the cells among a group of independent cells starts emitting an oscillating chemical signal of cyclic AMP, to attract other cells to join itself and form a temporal multi-cellular system, which subsequently differentiate into prestalk cells and prespore cells. Then, the prestalk cells drugs around the whole collection of cells and finally changes into a stalk, while the prespore calls clime up the stalk and change into spores at the top of the stalk, until the environmental condition recovers. They stay as a collection of single cells when the environmental condition is favored and they try to form multiple cell system when the environment is unfavored. Also this change is caused by transmission of chemical signal among cells. Assumption that the intercellular communication helped survival of monocellular society may be reasonable. This is again related to the failure of the thermodynamic condition FFE. The behavior of D. D. when the thermodynamic condition changed again into a favorite condition implies general tendency of evolution corresponding to the MEPP.

\subsection{Development of brain and Awareness}

Although the signal exchange system helped life in the unfavored condition, it would have finally helped activity of multi-cellular system and increased entropy production compared with a uni-cellular system without signal exchange, because this signal developed cooperation between the cells. Already in primitive multi-cellular structure such as hydra \cite{shimizu93}, the motion of tentacles are connected with a ling like neural network around the mouth. The neural network developed with evolution. At the late stage of evolution a concentrated neural system called brain was enhanced. Simultaneously, the sensor such as eyes were formed and the information of the sensor is sent to the brain and the information is processed for the use of the biosystem. Progress of brain function made human beings possible to model and predict the outside world. Tools thus invented are essential as the most specified characteristics for distinguishing human beings from other animals. The tools have been developed from the stone tools eventually to the developed information carriers and artificial intelligence.

Fire is a result of a chemical change from fuel materials with rich energy to materials of lower energy by a sort of chain reaction \cite{zevallos25}. The use of fire, the basis of modern industries, means use of this energy by controlling this chain reaction, and is directly related to the external entropy production. It is natural to imagine that tools which the ancient people had used for a long time had helped them to control the chain reaction of fire burning. It is only recently, however, that the technology has developed for creating gigantic products for variety of human use together with the technology of large amount of energy consumption needed for it. 

\section{Critical number of neurons for self-awareness necessary for invention of tools and 
control of fire}

Awareness or self-awareness has been a philosophical and psychological subject since long time ago, until recently when scientific research of these subjects started due to the development of brain research and computer science \cite{ishida04,lage22,kohda25,sawada12}. It is imagined that self-awareness is necessary for the invention of tool about 3.3 million years ago \cite{braidwood25}, as well as control of fire about $1.7-2.0$ million years ago \cite{james89}. 

\begin{figure}[t]
%\isPreprints{\centering}{} % Only used for preprints
\includegraphics[width=12.0 cm]{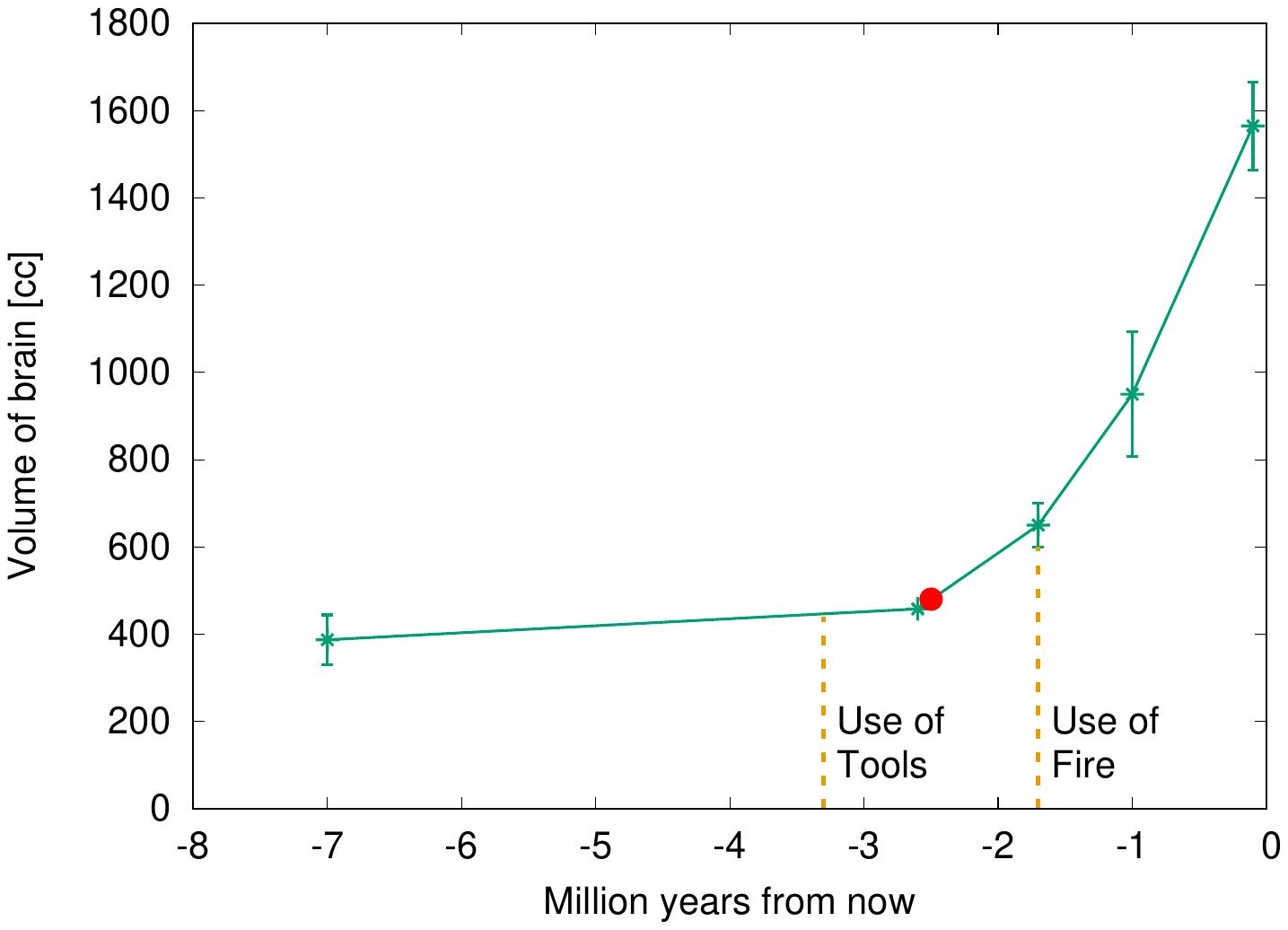}
\caption{
Archaeological data of brain size from chimpanzee to ancient human beings versus time. The green crosses correspond to the different types of homo with chimpanzee. The vertical dashed lines indicate the rough periods when the use of tools and control of fire started. The red point at $T_c=-2.5\;{\rm million~years}, N_c=480\;$cc is assumed as the critical point for the transition to a phase of new life with in a society and external entropy production.
\label{fig1}}
\end{figure}

History of appearance time of ancient human beings and their brain sizes after chimpanzee is summarized in Table~\ref{tab_brainsize}. The data of brain size from the references \cite{herndon99,neubauer12,tobias87,rightmire13,sherwood08} are plotted with the time of appearance in Figure~\ref{fig1}, which clearly shows that the brain size started increasing rapidly at a critical point $T_c=2.5$ million years ago where the brain size is about $N_c = 480\;$cc. This value may correspond to the historical time for human beings to have established self-awareness, because it corresponds to the period when usage of tools and controlling of fire started. Furthermore, this volume of brain corresponds roughly to the brain size of modern human baby at 1.5 years old, when it is believed that they obtain self-awareness \cite{holinger26}.

\begin{table}[h] 
%\small % Change table font size
\caption{History of appearance time of ancient human beings and their brain sizes after chimpanzee.\label{tab_brainsize}
}
\isPreprints{\centering}{} % Only used for preprints
\begin{tabularx}{\textwidth}{CCC}
\toprule
\textbf{Time}	& \textbf{Species}	& \textbf{Brain size}\\
\midrule
$\sim 7\;$million years ago		& Chimpanzee			& $387\pm57\;$cc \cite{herndon99}\\
$\sim 2.6\;$million years ago   & Australopithecus africanus & $458\pm6\;$cc \cite{neubauer12}   \\
$\sim 1.7\;$million years ago		& Homo habilis			& $650\pm50\;$cc \cite{tobias87}\\
$\sim 1\;$million years ago		& Homo erectus			& $950\pm143\;$cc \cite{rightmire13}\\
$\sim 0.1\;$million years ago	& Homo neanderthalensis 	& $1565\pm101\;$cc \cite{sherwood08}\\
\bottomrule
\end{tabularx}
%\noindent{\footnotesize{\textsuperscript{1} Tables may have a footer.}}
\end{table}

\unskip

Brain is a complicated network of neurons with synapses and its modeling mechanism of external world is a vast target of present researcher of brain science \cite{sherwood08}. Here, we discuss brain size as the simplest measure of brain function of ancient human beings.
%, although it is possible that the brain function may be expressed by a higher order of the brain volume.
It is generally accepted that prefrontal cortex is responsible for higher functions of human being. We used the total brain size $N$ instead of the volume of prefrontal cortex $N_p$, because the information of $N$ is archaeologically available, but not of $N_p$ for the ancient human beings. Therefore, when we assume that $N_c$ is $480\;$cc, it does not mean that the critical volume of $N_p$ for the birth of awareness is $480\;$cc. It only states that the critical volume of $N_p$ should corresponds to a value for which $N$ is $480\;$cc. The relation between $N_p$ and $N$ for the present young brains was obtained \cite{kanemura03,herculano05}, which would be useful for future construction of the model of birth of awareness for human baby. The legitimacy of using the same relation for the ancient human beings and for growing young brain is not clear at present.

By observing Figure~\ref{fig1}, together with the consideration in the previous section, it may be reasonable to make the following model for the birth of external entropy production for $N > N_c$:

(i) The ancient human beings around 2.5 million years ago started to cooperate with each other in trying to fabricate tools and trying to control fire, mainly due to increasing chance to meet each other due to gradual increase of population.

(ii) This cooperation for treating unaccustomed materials and phenomena accelerated awareness of human brain.

(iii) The accelerated brain helped growth of external world represented by the population of a group.

(iv) External entropy production was born as the result of the interaction between the internal world and external world by the awareness.

(v) Increase of population which uses fire means increase of external entropy production. 

\begin{figure}[t]
%\isPreprints{\centering}{} % Only used for preprints
\includegraphics[width=12.0 cm]{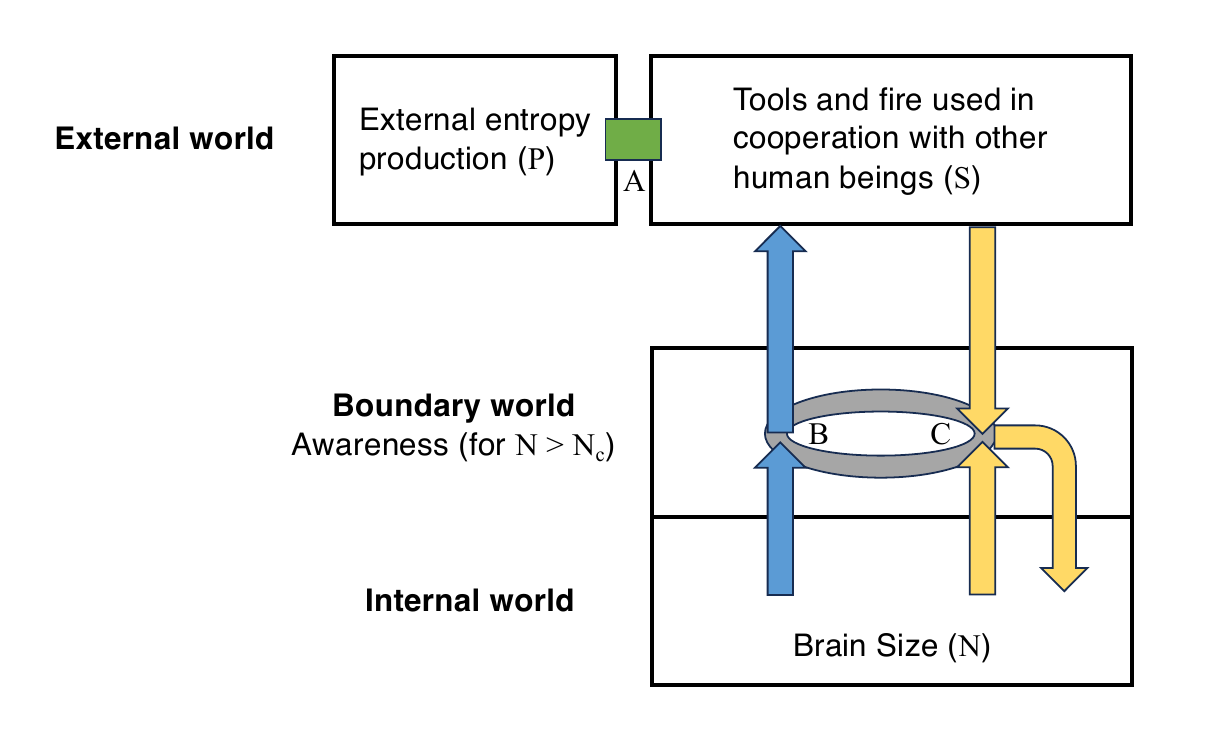}
\caption{
Interaction of internal and external worlds through awareness for $N > N_c$. The two worlds interacts through the boundary world representing awareness of human beings shown by the gray ring. The yellow arrows represent interaction of the external world and the brain through the awareness, described by Equation~(\ref{eq:dndt2}). The blue arrows represent action of the brain upon the external world through the awareness, described by Equation~(\ref{eq:dsdt}). The green bar represents the external entropy production by the external world described by Equation~(\ref{eq:entropy}). The internal world is quantitatively represented by brain size $N$. The external world represents tools and fire used in cooperation with other interacting human beings. The population size of a group $S$, which is the simplest quantity representing the external world, is used for present quantitative discussion. The external entropy production $P$ is assumed to be proportional to $S$ near the critical time, although this relation deviated from linearity very quickly.
\label{fig2}}
\end{figure}

In writing equations for this model, we use $N$, 
%brain size which includes the old part of brain plus the new part of brain responsible for the awareness or consciousness, $S$,
the effective size of external world $S$ which is represented by the population of a group to which the brain belongs, and 
%$P$, 
the amount of the external entropy production $P$. Diagram of the growth model of the brain size and the birth of a cooperating group which can produce external entropy production is shown in Figure~\ref{fig2}.

\unskip

%%%%%%%%%%%%%%%%%%%%%%%%%%%%%%%%%%%%%%%%%%
\section{Birth of external entropy production by a co-development of personal brain and external cooperative group}

\subsection{Coupled equations for development of personal brain with external world}

Although no mathematical analysis has existed so far \cite{gonzalez25,stibel25} to the best knowledge of the authors, Figure~\ref{fig1} provides us as an experimental data which shows qualitatively that brain size shows increasing roughly from the time of using tools and fire. Present model was shown from (i) to (v) in the previous section. Although the dependence of increasing speed of the brain size $N(t)$ on the interacting group size $S(t)$ is not known \cite{sherwood08}, it would be natural to assume that the brain growth late is function of personal brain size $N(t)$ and population size $S(t)$ in which the brain belongs. The diagram exhibiting the relation between the internal world represented numerically by brain size $N(t)$ and the external world represented numerically by interacting group size $S(t)$ through the awareness of the boundary world is shown in Figure~\ref{fig2}. 

The brain size $N(t)$ increases by $f(N,S)$, interaction of $N$ with external group size $S$, and decays with a natural decay $D$,
\begin{equation}
    \frac{dN(t)}{dt} = f[N(t), S(t)] - D[N(t)].
    \label{eq:dndt1}
\end{equation}
By putting as a simplest model,
\begin{equation}
    f[N(t), S(t)] = C N(t) S(t),~~~{\rm and}~~~ D[N(t)] = \gamma N(t),
    \label{eq:c}
\end{equation}
where $C$ is a coupling constant of $N$ and $S$, and $\gamma$ is a decay constant of $N$.
Equation~(\ref{eq:dndt1}) will be
\begin{equation}
    \frac{dN(t)}{dt} = CN(t)S(t) - \gamma N(t).
    \label{eq:dndt2}
\end{equation}

The size of an external world $S$ increases with the brain size $N$,
\begin{equation}
    \frac{dS(t)}{dt} = BN(t) - \mu S(t),
    \label{eq:dsdt}
\end{equation}
where $\mu$ is a decay constant of $S$.
The nonlinear equations (\ref{eq:dndt2}) and (\ref{eq:dsdt}) are a simplified coupled model of brain size $N$ and size of an external world $S$.

External entropy production per person $P$ will be proportional in the beginning to the newly developed external world $S$,
\begin{equation}
    P(t) = A S(t),
    \label{eq:entropy}
\end{equation}
where we set $A$ to be a constant.

It is noted that the detailed mechanism of the birth of awareness at $N = N_c$, nor $N$ dependence of the coupling constants $A, B, C$ at $N > N_c$ are not discussed here, and will be a subject of future study including brain science and evolutionary social science \cite{cavalli20}. 

\subsection{Linearized solution with numerical values obtained from the historical data}

To solve $N$ and $S$ by linearizing from Equations (\ref{eq:dndt2}) and (\ref{eq:dsdt}) as
\begin{equation}
    N = N_c + n(t), ~~~ S = S_c + s(t),
    \label{eq:linearize}
\end{equation}
where $N_c = \gamma\mu/BC$ and $S_c = \gamma/C$ are the equilibrium stage of $N-S$ space (i.e., $dN/dt = dS/dt = 0$). The red circle in Figure~\ref{fig1} corresponds to $N_c(T_c)$. Inserting Equation (\ref{eq:linearize}) into Equations (\ref{eq:dndt2}) and (\ref{eq:dsdt}), we obtain linearized equations for $n$ and $s$ as
\begin{equation}
    \frac{dn(t)}{dt} = CN_c s(t),
\end{equation}
\begin{equation}
    \frac{ds(t)}{dt} = Bn(t) - \mu s(t).
\end{equation}
By setting, $n=n(0)\exp(\lambda t)$ and $s=s(0)\exp(\lambda t)$, the eigenvalues $\lambda$ are found to be
\begin{equation}
    \lambda = \frac{1}{2}\left(-\mu \pm \sqrt{\mu^2 + 4 \gamma \mu}\right).
\end{equation}
Considering that $\mu \gg \gamma$, the mode with the positive eigenvalue is
\begin{equation}
    \lambda \approx \gamma.
\end{equation}
The growth solution is consistent with the MEPP.
Linear growth speed of brain coupled with the environmental grow at $N = N_c$ is determined by the decay speed of the brain. In order to obtain the information of numerical values utilized in this mode, we learn from Figure~\ref{fig1} as $N_c = 480\;$cc and
\begin{equation}
    \gamma n(0) = \frac{dN}{dt}(N = N_c) \sim 200\;{\rm cc}/{\rm million~years}.
\end{equation}
Assuming $n(0) \sim 10\;$cc as the fluctuation from $N_c$, the decay constant of brain obtained by this model is $\gamma \sim  20/$million years, which is not very far from 5/million year, a known decay constant of RNA in born \cite{allentoft12}. And assuming the critical group size $S_c \sim 10$ and the lifetime of a society $1/\mu \sim 10$ years, we estimate the coupling constants $C \sim 2/$million years and $B \sim  2 \times 10^3/{\rm cc} ~{\rm million ~years}$.

By the present analysis, the initiation of the external entropy production $P$ with formation of society of multi-body life was also confirmed as a manifestation of MEPP. Although the constant $A$ in equation (\ref{eq:entropy}) which relates the amount of entropy production $P$ and $S$ was not explicitly clarified in the present analysis, the growth rate is given by $\gamma \sim 20/$million years at $N = N_c$ and keep thereafter increasing exponentially. 
%Also, it is assumed in the present analysis that the coupling constant $C$ between the internal world $N$ and the external world $S$ is constant for $N > N_c$. Brain size dependence of awareness would be an important problem. Improvement of the model will be required for future study.
It should be mentioned here that internal entropy production inside brain also increases when $N > N_c$.

The coupling constant $C$ in Equation~(\ref{eq:c}) depends on the types of animals. The size of $C$ for human beings is considered much larger than those of other animals due the accuracy of fingers and sensitivity of eyes obtained by the life style. $N_c$ and $S_c$ are inversely proportional to $C$ as derived just below Equation~(\ref{eq:linearize}). This will suggest that the critical numbers of brain size and society size for human beings may be smaller than the other animals, and that this transition to the external entropy production is so far limited only to human beings.
Understanding the continued growth of the external entropy production is important, and for this purpose, future research of detailed nonlinear mode will be necessary in cooperation with the research of cultural evolution theory \cite{boyd88}.

%%%%%%%%%%%%%%%%%%%%%%%%%%%%%%%%%%%%%%%%%%
\section{Thermodynamic evolution theory of life and Neo-Darwinian theory}

The late time evolution investigated in the present paper were shown consistent with the birth and early time evolution described in the preceding paper with in the thermodynamic theory. Now, it may be worth mentioning about the relation between the thermodynamic evolution theory and Neo-Darwinian theory. We focus here on the targets and time scales of the two theories.

\subsection{Neo-Darwinian evolution theory}

Darwin expressed evolution of life as the survival of the fittest \cite{darwin59}.  After Mendel’s discovery of the rule for inheritance over generation, Neo-Darwinism \cite{hancock21}, appeared as a modified theory of evolution, taking into account the gene information and its mutation. Neo-Darwinian theory is concerned with diversity of species and natural selection among them qualitatively and concerned with natural selection by fitness to the environment. The cause of changes is due to the mutation of genes, with possible epigenetic modification of DNA strand \cite{liu21}. The former is known as neutral, and the latter is now being investigated whether the effect is maintained over the generations. Neo-Darwinian theory works with the environmental change and the fitness of any part of biological system at the time interval of interest. The selection in evolution is done by fitness to the environment. To predict change the information of environmental change will be necessary.

\subsection{Thermodynamic evolution theory}

This theory has its basis on the second law of thermodynamics. This viewpoint insists that the birth and successive evolution should consistently be described by the MEPP, when the local system under consideration is FFE and free for pumping of the produced entropy. The time scale is long because the thermodynamics is fundamentally stochastic. Target of MEPP is for a local system FFE in an infinitely large reservoir, and time scale of MEPP is also long if the environment of the local system stays unchanged. The selection rule in evolution is limited to the amount of entropy production no matter what kind of change occurs in the environment.

\subsection{Comparison of the two theories}

Thermodynamic theory discusses entropy production only over a long time scale, while fitness of the Neo-Darwinian theory is more restricted to short time scale because fitness is a concept of short time scale. Target of Neo-Darwinian theory covers any features of the living bodies. The theory works fine if the fitness can be found at the time interval of interest. 

Because the target is free, it may not be easy to discuss the fitness of the unknown future. On the other hand, target quantity of thermodynamic theory of evolution in only entropy production. The applicability is limited by the condition of the local system under consideration to be open and far from equilibrium. The advantage of thermodynamic evolution theory is ability to treat the birth and future of life consistently.

%%%%%%%%%%%%%%%%%%%%%%%%%%%%%%%%%%%%%%%%%%
\section{Human being as a symbiotic existence of multi-cellular and multi-body lives}

Important birth of awareness in human brain was initially to avoid dangers from environment, but after a while it was used for making tools and controlling fire and initiate technology with a large amount of external entropy production, which is in accordance with the MEPP. The result of present research implied that production of external entropy production is nothing but a birth of the second life, as much as the birth of the first life which started with sudden increase of internal entropy production \cite{sawada25}. This situation reminds us of D.D. which switches between a single cell state to multi-cell state depending on the environmental condition as described in Section 2-2. As a result of evolution to external entropy production , life of human being bears two aspects; multi-cellular life and a multi-body live. The former has a long history like $\sim 2$ billion years, and the latter social life has a relatively short history like $\sim 2$ million years. This symbiotic structure has creates some popular features in the behavior of human beings.

\subsection{Coexistence of two kinds of entropy production mechanisms in a body}

Human society keeps increasing the amount of external entropy production, and at the same time individual human being keeps desiring for individual survival. These two mechanisms function by two different time scales and at two different locations in a body. Therefore, the behavior of human being reflects these two functions. It is known that consciousness of homo-sapiens reflects the function of frontal cortex of brain, because it is accompanied by awareness. The functions of other part of brain as well as of the body part which are carried out without consciousness and appears only at some unconscious time. This almost hidden factor may correspond to the phenomena called instinct \cite{blumberg16}. Therefore, thermodynamic evolution theory may contribute to the researches in the field of psychology, ethology and evolutional psychology. 

\subsection{Why technology grows continuously?}

Human being keeps promoting technology without knowing the goal. From a long time scale consideration on the external entropy production, human technology was fundamentally introduced by the thermodynamic principle, more precisely speaking by MEPP when the system is FFE. Therefore, the large amount of entropy production which modern technology produces is not necessarily biproduct of the convenience for human being. On the contrary, the large amount of entropy production itself is, scientifically speaking, the goal of nature in FFE condition, and the tools and technology is only the medium for the external entropy production. This reversal of the cause and result may help obtaining an answer for a question why present human beings continue developing technology with increasing entropy production.

%%%%%%%%%%%%%%%%%%%%%%%%%%%%%%%%%%%%%%%%%%
\section{Conclusion}

A new concept "external entropy production" proposed recently in thermodynamic evolution theory, in contrast to "internal entropy production" inside the living body is investigated theoretically. It is pointed out from the archaeological data that growth of human brain with birth of awareness were indispensable for the invention of tools and control of fire, which had triggered interaction with external world. Brain size as the simplest measure of brain function and the interacting group size are useful for a mathematical treatment for the birth of awareness and external entropy production. Birth of external entropy production is theoretically formalized by coupled equations of the personal brain size and the interacting population size through awareness. The theoretical results are compared with the data of the archaeological findings. The present analysis of the beginning of the external entropy production implies that production of external entropy production is nothing but a birth of the second life, as much as the birth of the first life which started sudden increase of internal entropy production. Long time viewpoint of thermodynamic evolution suggests us of a new viewpoint on the meaning of technology and self-awareness which have been interpreted only short time scale less than a few hundred years. This finding may give us a new impact from philosophical viewpoint. It is suggested that a symbiotic existence of human being with personal internal MEPP and external MEPP may be deeply related with a psychological phenomena of instinct. It is hoped that the thermodynamic understanding of the external entropy production may contribute to build countermeasure against global warming for future human life.

%%%%%%%%%%%%%%%%%%%%%%%%%%%%%%%%%%%%%%%%%%
%\section{Patents}

%This section is not mandatory, but may be added if there are patents resulting from the work reported in this manuscript.

%%%%%%%%%%%%%%%%%%%%%%%%%%%%%%%%%%%%%%%%%%
\vspace{6pt} 

%%%%%%%%%%%%%%%%%%%%%%%%%%%%%%%%%%%%%%%%%%
%% optional
%\supplementary{The following supporting information can be downloaded at:  \linksupplementary{s1}, Figure S1: title; Table S1: title; Video S1: title.}


% Only used for preprtints:
% \supplementary{The following supporting information can be downloaded at the website of this paper posted on \href{https://www.preprints.org/}{Preprints.org}.}

%\begin{tabularx}{\textwidth}{lll}
%\toprule
%\textbf{Name} & \textbf{Type} & \textbf{Description} \\
%\midrule
%S1 & Python script (.py) & Script of python source code used in XX \\
%S2 & Text (.txt) & Script of modelling code used to make Figure X \\
%S3 & Text (.txt) & Raw data from experiment X \\
%S4 & Video (.mp4) & Video demonstrating the hardware in use \\
%\bottomrule
%\end{tabularx}
%}

%%%%%%%%%%%%%%%%%%%%%%%%%%%%%%%%%%%%%%%%%%
\authorcontributions{Conceptualization, Y.S.; methodology, Y.S. and K.T.; validation, Y.S., and K.T.; writing—original draft preparation, Y.S.; writing—review and editing, Y.S. and K.T.; visualization, K.T. All authors have read and agreed to the published version of the manuscript.}

\funding{This research received no external funding.}

\institutionalreview{Not applicable.}

\informedconsent{Not applicable.}

\dataavailability{Not applicable.}

\conflictsofinterest{The authors declare that the research was conducted in the absence of any commercial or financial relationships that could be construed as a potential conflict of interest.} 

%%%%%%%%%%%%%%%%%%%%%%%%%%%%%%%%%%%%%%%%%%
%% Optional

%% Only for journal Encyclopedia
%\entrylink{The Link to this entry published on the encyclopedia platform.}

%\abbreviations{Abbreviations}{
%The following abbreviations are used in this manuscript:
%\\

%\noindent 
%\begin{tabular}{@{}ll}
%MDPI & Multidisciplinary Digital Publishing Institute\\
%DOAJ & Directory of open access journals\\
%TLA & Three letter acronym\\
%LD & Linear dichroism
%\end{tabular}
%}

%%%%%%%%%%%%%%%%%%%%%%%%%%%%%%%%%%%%%%%%%%
\isPreprints{}{% This command is only used for ``preprints''.
\begin{adjustwidth}{-\extralength}{0cm}
} % If the paper is ``preprints'', please uncomment this parenthesis.

\reftitle{References}


% \bibliography{your_external_BibTeX_file}

\begin{thebibliography}{999}
\bibitem{sawada25}
Sawada, Y.; Daigaku, Y.; Toma, K. Maximum Entropy Production Principle of Thermodynamics for the Birth and Evolution of Life. {\em Entropy} {\bf 2025}, {\em 27}, 449.
\bibitem{zeraati12}
Zeraati, S.; Jafarpour, F.H.; Hinrichsen, H. Entropy production of nonequilibrium steady states with irreversible transitions. {\em J. Stat. Mech.} {\bf 2012} L12001.
\bibitem{martyushev21}
Martyushev, L.M. Maximum entropy production principle: History and current status. {\em Phys. Uspekhi} {\bf 2021}, {\em 64}, 558–583.
\bibitem{england13}
England, J.L. Statistical physics of self-replication {\em J. Chem. Phys.} {\bf 2013}, {\em 139}, 121923.
\bibitem{malkus58}
Malkus, W.V.R.; Veronis, G. Finite Amplitude Cellular Convection. {\em J. Fluid. Mech.} {\bf 1958}, {\em 4}, 225–260.
\bibitem{ben90}
Ben-Jacob, E.;. Garik, P. The formation of patterns in non-equilibrium  growth. {\em Nature} {\bf 1990}, {\em 343}, 523–530.
\bibitem{hill90}
Hill, A. Entropy production as the selection rule between different growth morphologies. {\em Nature} {\bf 1990}, {\em 348}, 426–428.
\bibitem{arendt21}
Arendt, D. Elementary Nervous Systems. {\em Philos. Trans. R. Soc. B Biol. Sci.} {\bf 2021}, {\em 376}, 20200347. 
\bibitem{glansdorff71}
Glansdorff, P.; Prigogine, I. Structure, {\em Stability and Fluctuations.}; Masson et Cie: Echandens, Switzerland, 1971.
\bibitem{dusenberry96}
Dusenberry, D.B. {\em Life at Small Scale: The Behavior of Microbes.}; Holt \& Company, Henry, 1996.
\bibitem{shimizu93}
Shimizu, H.; Sawada, Y.; Sugiyama, T. Minimum tissue size required for hydra regeneration. {\em Dev. Biol.} {\bf 1993} {\em 155}, 287-296.
\bibitem{zevallos25}
Zevalles, A.A.M. {\em Flame Propagation Fundamentals: Applications to Sustainable Fuels}; Springer International Publishing AG, 2025.
\bibitem{ishida04}
Ishida1, F.; Sawada, Y. Human Hand Moves Proactively to the External Stimulus: An Evolutional Strategy for Minimizing Transient Error. {\em Phys. Rev. Lett.} {\bf 2004} {\em 93}, 168105.
\bibitem{lage22}
Lage, C.A.; Wolmarans, D.W.; Mograbi, D.C. An evolutionary view of self-awareness. {\em Behav. Processes}, {\bf 2022} {\em 194}, 104543.
\bibitem{kohda25}
Kohda, M.; Sogawa, S.; Bshary, R. On the mirror test and the evolutionary origin of self-awareness in vertebrates. {\em Philos. Trans. R. Soc. B Biol. Sci.} {\bf 2025}, {\em 380}, 20240312.
\bibitem{sawada12}
Sawada, Y. The Aspects, the Origin, and the Merit of Aware Computing. {\em Applied Computational Intelligence and Soft Computing} {\bf 2012} 760908.
\bibitem{braidwood25}
Braidwood, R.J. Stone Age. {\em Britannica anthropology} {\bf 2026}.
\bibitem{james89}
James, S.R. Hominid Use of Fire in the Lowe and Middle Pleistocene: A Review of the Evidence. {\em Current Anthropology} {\bf 1989} {\em 30}, 1–26.
\bibitem{herndon99}
Herndon, J.G.; Tigges, J.; Anderson, D.C.; Klumpp, S.A.; McClure, H.M. Brain weight throughout the life span of the chimpanzee. {\em J. Comp. Neurol.} {\bf 1999} {\em 409}, 567-72.
\bibitem{neubauer12}
Neubauer, S.; Gunz, P.; Webe, G.W.; Hublin, J.-J. Endocranial volume of Australopithecus africanus: New CT-based estimates and the effects of missing data and small sample size. {\em J. Hum. Evol.} {\bf 2012} {\em 62}, 498-510.
\bibitem{tobias87}
Tobias, P.V. The brain of Homo habilis: A new level of organization in cerebral evolution. {\em J. Hum. Evol.} {\bf 1987} {\em 16}, 741-761.
\bibitem{rightmire13}
Rightmire, G.P. Homo erectus and Middle Pleistocene hominins: Brain size, skull form, and species recognition. {\em J. Hum. Evol.} {\bf 2013} {\em 65}, 223-52.
\bibitem{sherwood08}
Sherwood, C.C.; Subiaul, F.; Zawidzki, T.W. A natural history of the human mind: tracing evolutionary changes in brain and cognition. {\em J. Anat.} {\bf 2008} {\em 212}, 426-454.
\bibitem{holinger26}
Holinger, P.C. How Self-Awareness Starts: Self-awareness has its roots in the transition from infant to toddler. {\em Psychology Today.} {\bf 2026}, https://www.psychologytoday.com.
\bibitem{kanemura03}
Kanemura, H.; Aihara, M.; Aoki, S.; Araki, T.; Nakazawa, S. Development of the prefrontal lobe in infants and children: a three-dimensional magnetic resonance volumetric study. {\em Brain Dev.} {\bf 2003} {\em 25}, 195-9.
\bibitem{herculano05}
Herculano-Houzel, S.; Lent, R. Isotropic Fractionator: A Simple, Rapid Method for the Quantification of Total Cell and Neuron Numbers in the Brain. {\em J. Neurosci.} {\bf 25}, 2518–2521.
\bibitem{gonzalez25}
Gonz\'{a}lez-Forero, M.; G\'{o}mez-Robles,a A. Why did the human brain size evolve? A way forward. {\em Philos. Trans. R. Soc. B Biol. Sci.} {\bf 2025}, {\em 380}, 20240114.
\bibitem{stibel25}
Stibel, J.M. Did increasing brain size place early humans at risk of extinction? {\em Brain Cogn.} {\bf 2025} {\em 188}, 106336.
\bibitem{cavalli20}
Cavalli-Sforza, L.L.; Feldman, M.W. {\em Cultural Transmission and Evolution: A quantitative approach}; Princeton University Press, 2020.
\bibitem{allentoft12}
Allentoft, M.E.; Collins, M.; Harker, D.; et al. The half-life of DNA in bone: measuring decay kinetics in 158 dated fossils. {\em Proc. Biol. Sci.} {\bf 2012} {\em 279}, 4724-4733.
\bibitem{boyd88}
Boyd, R.; Richerson, P.J. {\em Culture and the Evolutionary Process}; University of Chicago Press, 1988.
\bibitem{darwin59}
Darwin, C. {\em On the Origin of Species by Means of Natural Selection, or the Preservation of Favoured Races in the Struggle for Life}; John Murray: London, UK, 1859.
\bibitem{hancock21}
Hancock, Z.B.; Lehmberg, E.S.; Bradburd, G.S. Neo-Darwinism still haunts evolutionary theory: A modern perspective on Charlesworth, Lande, and Slatkin (1982). {\em Evolution} {\bf 2021} {\em 75}, 1244-1255.
\bibitem{liu21}
Liu, J.; Mosti, F.; Silver, D.L. Human brain evolution: Emerging roles for regulatory DNA and RNA. {\em Crr. Opin. Neurobiol.} {\bf 2021} {\em 71}, 170-177.
\bibitem{blumberg16}
Blumberg, M.S. Development evolving: The origins and meanings of instinct. {\em Wiley Interdiscip. Rev. Cogn. Sci.} {\bf 2016} {\em 8}, 10.1002.

\end{thebibliography}

%=====================================
% References, variant B: internal bibliography
%=====================================

% ACS format
\isAPAandChicago{}{%

}

%%%%%%%%%%%%%%%%%%%%%%%%%%%%%%%%%%%%%%%%%%
%% for journal Sci
%\reviewreports{\\
%Reviewer 1 comments and authors’ response\\
%Reviewer 2 comments and authors’ response\\
%Reviewer 3 comments and authors’ response
%}
%%%%%%%%%%%%%%%%%%%%%%%%%%%%%%%%%%%%%%%%%%
\PublishersNote{}
\isPreprints{}{% This command is only used for ``preprints''.
\end{adjustwidth}
} % If the paper is ``preprints'', please uncomment this parenthesis.
\end{document}